\documentclass[twocolumn,aps,prl,showpacs]{revtex4}
\usepackage[T1]{fontenc}
\usepackage[latin9]{inputenc}
\usepackage{amsmath}
\usepackage{amssymb}
\usepackage{graphicx}
\usepackage{esint}

\makeatletter
\@ifundefined{textcolor}{}
{%
 \definecolor{BLACK}{gray}{0}
 \definecolor{WHITE}{gray}{1}
 \definecolor{RED}{rgb}{1,0,0}
 \definecolor{GREEN}{rgb}{0,1,0}
 \definecolor{BLUE}{rgb}{0,0,1}
 \definecolor{CYAN}{cmyk}{1,0,0,0}
 \definecolor{MAGENTA}{cmyk}{0,1,0,0}
 \definecolor{YELLOW}{cmyk}{0,0,1,0}
}

\makeatother

\begin{document}

\title{Gapless topological Fulde-Ferrell superfluidity induced by in-plane
Zeeman field}

\author{Hui Hu$^{1}$, Lin Dong$^{2}$, Ye Cao$^{1}$, Han Pu$^{2}$, and
Xia-Ji Liu$^{1}$}

\affiliation{$^{1}$Centre for Quantum and Optical Science, Swinburne University
of Technology, Melbourne 3122, Australia}

\affiliation{$^{2}$Department of Physics and Astronomy, and Rice Quantum Institute,
Rice University, Houston, TX 77251, USA}

\date{\today}
\begin{abstract}
\textbf{Topological superfluids are recently discovered quantum matters
that host topologically protected gapless edge states known as Majorana
fermions - exotic quantum particles that act as their own anti-particles
and obey non-Abelian statistics. Their realizations are believed to
lie at the heart of future technologies such as fault-tolerant quantum
computation. To date, the most efficient scheme to create topological
superfluids and Majorana fermions is based on the Sau-Lutchyn-Tewari-Das
Sarma model with a Rashba-type spin-orbit coupling on the }\textbf{\textit{x-y}}\textbf{
plane and a large out-of-plane (perpendicular) Zeeman field along
the }\textbf{\textit{z}}\textbf{-direction. Here we propose an alternative
setup, where the topological superfluid phase is driven by applying
an in-plane Zeeman field. This scheme offers a number of new features,
notably Cooper pairings at finite centre-of-mass momentum (i.e., Fulde-Ferrell
pairing) and gapless excitations in the bulk. As a result, a novel
gapless topological quantum matter with inhomogeneous pairing order
parameter appears. It features unidirected Majorana surface states
at boundaries, which propagate in the same direction and connect two
Weyl nodes in the bulk. We demonstrate the emergence of such an exotic
topological matter and the associated Majorana fermions in spin-orbit
coupled atomic Fermi gases and determine its parameter space. The
implementation of our scheme in semiconductor/superconductor heterostructures
is briefly discussed.}
\end{abstract}

\pacs{05.30.Fk, 03.75.Hh, 03.75.Ss, 67.85.-d}

\maketitle
The possibility of realizing topological superfluids and manipulating
Majorana fermions in solid-state and ultracold atomic systems is currently
a topic of great theoretical and experimental interest \cite{Hasan2010,Qi2011},
due to their potential applications in fault-tolerant topological
quantum computation \cite{Nayak2008}. Roughly, Majorana fermions
constitute ``half'' of an ordinary Dirac fermion, in the sense that
two real Majorana fermions $\gamma_{1}$ and $\gamma_{2}$ - which
can be separated in arbitrary distance - mathematically define a complex
fermion operator $c=\gamma_{1}+i\gamma_{2}$ \cite{Wilczek2009}.
The exchange statistics of Majorana fermions is exotic. Unlike conventional
bosons and fermions, braiding Majorana fermions around one another
in a $2^{N}$-dimensional Hilbert space (spanned by $2N$ well-separated
Majorana fermions) produce non-Abelian unitary transformations in
the Hilbert space. Quantum information can then be non-locally encoded
in the Hilbert space by such braiding operators and be immune to decoherence,
which is ideal for the purpose of quantum computation \cite{Nayak2008}.

At present, a number of experimental setting have been suggested for
hosting Majorana fermions under appropriate conditions, including
chiral \textit{p}-wave superconductors \cite{Read2000}, fractional
quantum Hall systems at filling $\nu=5/2$ \cite{Moore1991}, and
topological insulators or semiconductor nanowires in proximity to
an \textit{s}-wave superconductor \cite{Fu2008,Sau2010,Oreg2010}.
The latest setting, which seems to be the most practical setup, is
described by the Sau-Lutchyn-Tewari-Das Sarma (SLTD) model \cite{Sau2010}.
The key idea of this mechanism is that the Fermi surfaces are spin-split
by a Rashba spin-orbit coupling on the $x$-$y$ plane and a perpendicular
out-of-plane Zeeman field along the $z$-direction. If the number
of particles is tuned to make the inner Fermi surface disappear, the
superconductivity will be only induced by pairing on the outer Fermi
surface, which is of \textit{p}-wave in nature \cite{Zhang2008,Sato2009},
and therefore becomes topologically nontrivial. Following this promising
theoretical model, exciting experimental progress for the observation
of Majorana fermions has been made very recently \cite{Mourik2012,Williams2012,Rokhinson2012},
although unambiguous confirmation for their existence still remains
elusive.

\begin{figure}
\begin{centering}
\includegraphics[clip,width=0.48\textwidth]{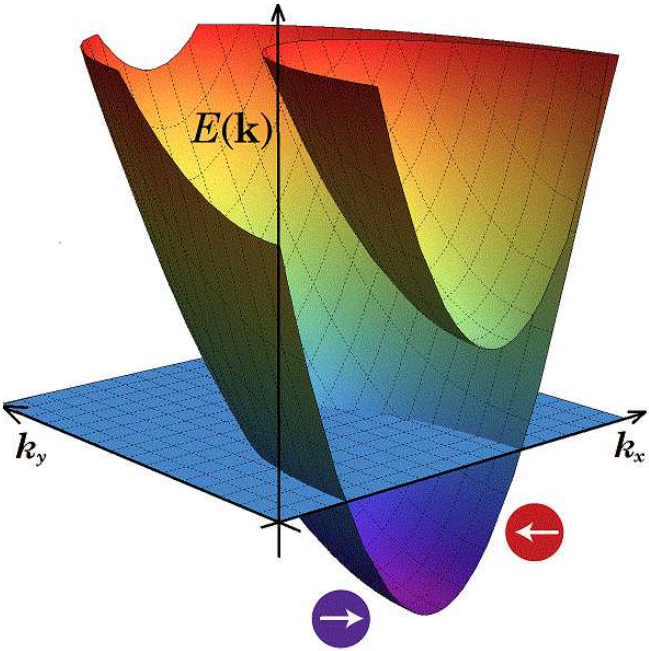} 
\par\end{centering}

\caption{(Color online). \textbf{Formation of an effective }\textbf{\textit{p}}\textbf{-wave
energy band}. The energy band of the system splits into two by spin-orbit
coupling. A large in-plane Zeeman field along the \textit{x}-direction
strongly tilts the energy dispersion. When the Fermi energy lies below
the upper band, atoms occupy the lower band only and form an effective
``spinless'' system, in which the new composite particle consists
of both spin-up and spin-down ingredients (shown by circles with arrows)
and interacts with each other via an effective \textit{p}-wave interaction.}

\label{fig1} 
\end{figure}

The SLTD mechanism uses an out-of-plane Zeeman field to split the
Fermi surfaces \cite{Sau2010}. It is interesting that such a splitting
can also be achieved by applying a large in-plane\emph{ }Zeeman field
in combination with spin-orbit coupling, as illustrated in Fig. \ref{fig1}.
Furthermore, the in-plane field together with spin-orbit coupling
is known to introduce an asymmetry in the single-particle dispersion
\cite{Dong2013PRA,Shenoy2013}, consequently to induce Cooper pairs
with nonzero centre-of-mass momentum \cite{Zheng2013,Wu2013,Liu2013,Dong2013NJP,Hu2013,Qu2013,Zhang2013,Fai2014}
and hence realize the so-called spatially inhomogeneous Fulde-Ferrell
(FF) pairing scenario \cite{Fulde1964}. It is therefore of interest
to ask whether a topological phase transition can also be driven by
an in-plane Zeeman field only? If the answer is yes, then we must
be able to observe an exotic inhomogeneous topological FF superfluid
that supports Majorana fermions. The understanding of such a new-type
topological state of matter may greatly enrich our knowledge about
topological superfluids.

In this work, we examine the new mechanism by using an ultracold atomic
setting of a three-dimensional (3D) spin-orbit coupled atomic Fermi
gas subject to an in-plane Zeeman field. By increasing the field strength
above a threshold, we observe the change of the topology of the Fermi
surfaces that triggers a topological phase transition. The resulting
inhomogeneous topological FF superfluid is gapless in the bulk with
nodal points form closed surfaces in momentum space. While in real
space, it hosts unidirected Majorana surface states that propagate
in the same direction at boundary. These unique features are absent
in standard topological superfluids known so far such as the \textit{p}-wave
superconductors or the SLTD-type superconductors, both of which are
gapped in the bulk and support counter-propagating Majorana modes
at surfaces. We find that the phase space for the proposed novel gapless
topological FF superfluid is significant, implying that it could be
easily realized in current ultracold atomic experiments owing to the
unprecedented tunability of synthetic spin-orbit coupling, Zeeman
field and interatomic interaction in cold-atom laboratory \cite{Lin2011,Wang2012,Cheuk2012}.
We also discuss briefly the potential implementation of our proposal
in solid-state setups.

\begin{figure}
\begin{centering}
\includegraphics[clip,width=0.48\textwidth]{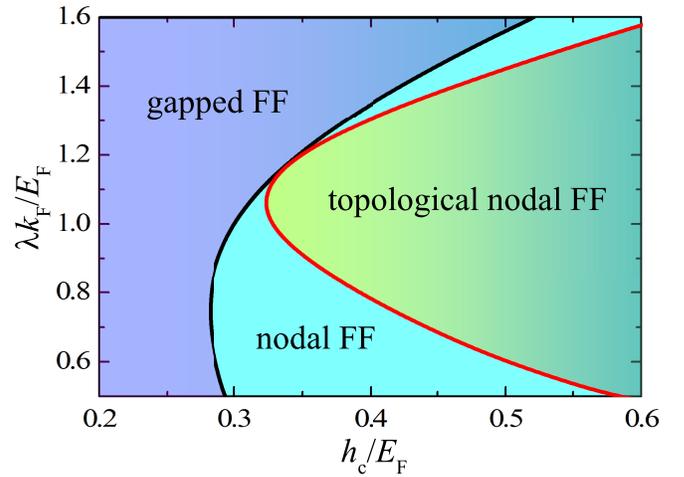} 
\par\end{centering}

\caption{(Color online) \textbf{Zero temperature phase diagram of the FF superfluid
at the interaction parameter $1/(k_{F}a_{s})=-0.5$}. With increasing
the in-plane Zeeman field, the Fermi cloud changes from a gapped FF
superfluid to a gapless FF superfluid, and finally turns into a gapless
topological superfluid. }

\label{fig2} 
\end{figure}

\textbf{\textit{Results}}. --- For concreteness, we focus on a 3D
spin-orbit coupled two-component Fermi gas with an isotropic spin-orbit
coupling $V_{SO}(\mathbf{\hat{k}})=\lambda(\hat{k}_{x}\sigma_{x}+\hat{k}_{y}\sigma_{y}+\hat{k}_{z}\sigma_{z})$
subject to an in-plane Zeeman field $h\sigma_{x}$ \cite{Dong2013NJP,Anderson2012,Anderson2013},
which can be described by the model Hamiltonian, 
\begin{equation}
\mathcal{H}=\int d{\bf r}\left[\sum_{\sigma\sigma'}\mathcal{\psi_{\sigma}^{\dagger}\left(\mathbf{r}\right)\textrm{\ensuremath{H}}}_{0}^{\sigma\sigma'}\psi_{\sigma'}\left(\mathbf{r}\right)+\mathcal{V}_{int}\right],
\end{equation}
where $\psi_{\sigma}^{\dagger}({\bf r})$ ($\psi_{\sigma}$) is the
field operator for creating (annihilating) an atom with pseudo-spin
state $\sigma\in(\uparrow,\downarrow)$ at position ${\bf r}$, $H_{0}=-\hbar^{2}\nabla^{2}/(2m)-\mu+V_{SO}(\mathbf{\hat{k}})+h\sigma_{x}$
is the single-particle Hamiltonian with the atomic mass $m$ and chemical
potential $\mu$, $\hat{k}_{i=(x,y,z)}=-i\partial_{i}$ is the momentum
operator and $\sigma_{x,y,z}$ are the Pauli matrices. $\text{\ensuremath{\mathcal{V}}}_{int}=U_{0}\psi_{\uparrow}^{\dagger}({\bf r})\psi_{\downarrow}^{\dagger}({\bf r})\psi_{\downarrow}({\bf r})\psi_{\uparrow}({\bf r})$
describes a pairwise attractive contact interaction of strength $U_{0}<0$,
where $U_{0}^{-1}=m/(4\pi\hbar^{2}a_{s})-V^{-1}\sum_{\mathbf{k}}m/(\hbar^{2}\mathbf{k}^{2})$
can be expressed in terms of the \textit{s}-wave scattering length
$a_{s}$. At the mean-field level, the model Hamiltonian can be solved
by taking an order parameter $\Delta({\bf r})=-U_{0}\left\langle \psi_{\downarrow}(\mathbf{{\bf r}})\psi_{\uparrow}(\mathbf{{\bf r}})\right\rangle $
and linearizing the interaction Hamiltonian ${\cal V}_{int}\simeq-[\Delta(\mathbf{{\bf r}})\psi_{\uparrow}^{\dagger}(\mathbf{{\bf r}})\psi_{\downarrow}^{\dagger}(\mathbf{{\bf r}})+\textrm{H.c.}]-\left|\Delta({\bf r})\right|^{2}/U_{0}$.

\begin{figure*}
\begin{centering}
\includegraphics[clip,width=0.9\textwidth]{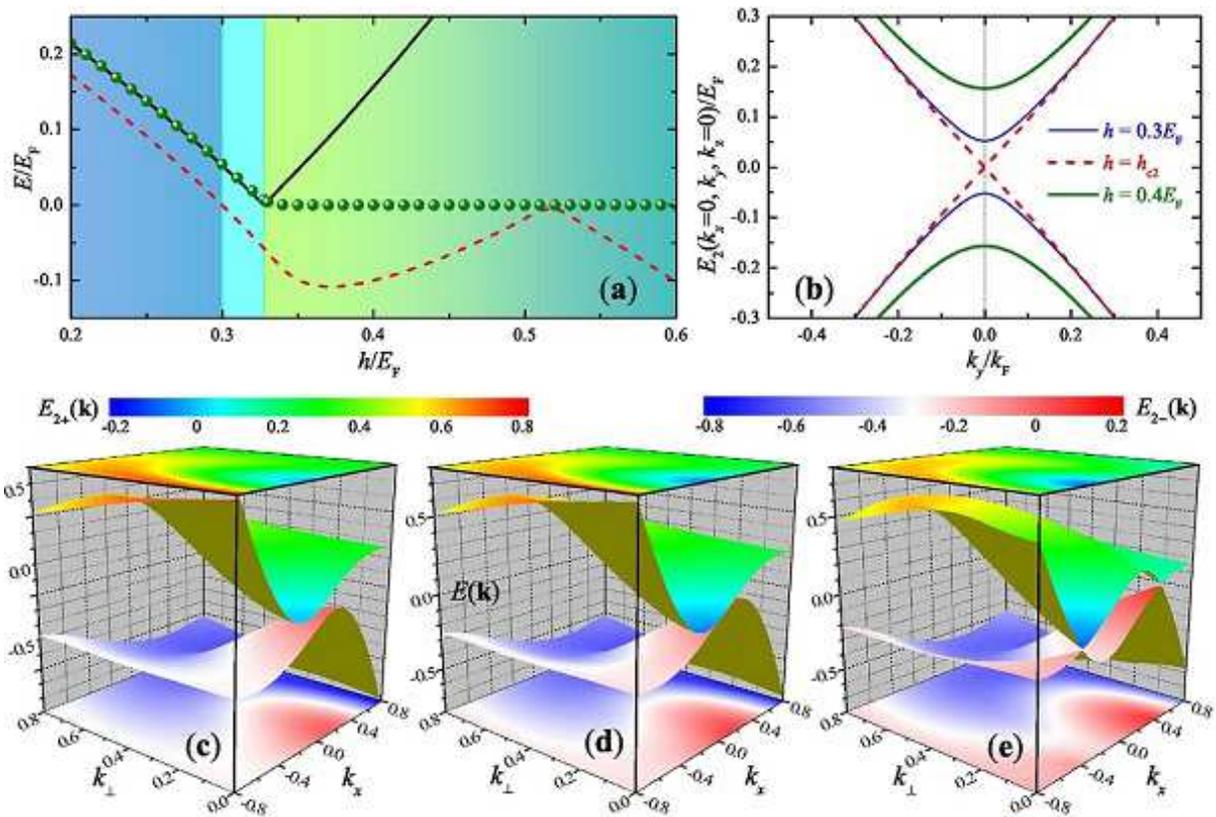} 
\par\end{centering}

\caption{(Color online) \textbf{The evolution of the energy gap and of the
topology of the Fermi surfaces at $\lambda=E_{F}/k_{F}$ with increasing
in-plane Zeeman field}. (\textbf{a}) The global energy gap $E_{g}=\min E_{2+}(\mathbf{k})$
(red dashed line), the energy gap at $\mathbf{k}=0$ (black solid
line), and the minimum energy of the surface states (green solid circles)
when open boundary is imposed to the \textit{y}-\textit{z} plane.
(\textbf{b}) The energy dispersion $E_{2\pm}(k_{y})$ at $k_{x}=0$
and $k_{z}=0$. (\textbf{c}), (\textbf{d}) and (\textbf{e}) The 3D
full plot of the energy dispersion $E_{2\pm}(k_{x},k_{\perp}=\sqrt{k_{y}^{2}+k_{z}^{2}})$
at $h_{c1}\simeq0.3E_{F}$ (c), $h_{c2}\simeq0.327E_{F}$ (d) and
$h=0.4E_{F}$ (e).}

\label{fig3} 
\end{figure*}

In the presence of an in-plane Zeeman field $h\sigma_{x}$, it is
now widely understood that Cooper pairs acquire a finite centre-of-mass
momentum $\mathbf{Q}=q\mathbf{e}_{x}$ along the \textit{x}-direction,
i.e., $\Delta({\bf r})=\Delta e^{iqx}$ \cite{Dong2013PRA,Zheng2013,Wu2013,Liu2013,Dong2013NJP}.
By using the Nambu spinor $\Phi(\mathbf{{\bf r}})\equiv[\psi_{\uparrow}e^{+iqx/2},\psi_{\downarrow}e^{+iqx/2},\psi_{\uparrow}^{\dagger}e^{-iqx/2},\psi_{\downarrow}^{\dagger}e^{-iqx/2}]$$^{T}$
to gauge out the momentum related phase in the order parameter, the
mean-field model Hamiltonian can be solved by diagonalizing the following
Bogoliubov-de Gennes (BdG) Hamiltonian

\begin{equation}
\mathcal{H}_{BdG}(\mathbf{\hat{k}})\equiv\left[\begin{array}{cc}
H_{0}\left(\frac{\mathbf{Q}}{2}+\mathbf{\hat{k}}\right) & -i\Delta\sigma_{y}\\
i\Delta\sigma_{y} & -H_{0}^{*}\left(\frac{\mathbf{Q}}{2}-\mathbf{\hat{k}}\right)
\end{array}\right],\label{eq:BdGHami}
\end{equation}
i.e., $\mathcal{H}_{BdG}\Phi_{\mathbf{k\eta}}^{\nu}(\mathbf{{\bf r}})=E_{\mathbf{\eta\nu}}(\mathbf{k})\Phi_{\mathbf{k\eta}}^{\nu}(\mathbf{{\bf r}})$,
which gives rise to the wavefunction of Bogoliubov quasiparticles,
$\Phi_{\mathbf{k\eta}}^{\text{\ensuremath{\nu}}}(\mathbf{{\bf r}})=1/\sqrt{V}e^{i\mathbf{k\cdot}\mathbf{{\bf r}}}[u_{\mathbf{\eta\uparrow}}^{\nu},u_{\mathbf{\eta\downarrow}}^{\nu},v_{\mathbf{\eta\uparrow}}^{\nu},v_{\eta\downarrow}^{\nu}]^{T}$,
and the energy $E_{\eta\nu}(\mathbf{k})$. We obtain four quasi-particle
energy dispersions, indexed by $\nu\in(+,-)$ for the particle ($+$)
or hole ($-$) branch, and $\text{\ensuremath{\eta\in}(1,2)}$ for
the upper ($1$) or lower ($2$) helicity band split by spin-orbit
coupling. We derive the gap and number equations from the resulting
mean-field thermodynamic potential (see Methods) and solve them self-consistently
to obtain $\Delta$, $q$ and $\mu$, from which we determine the
phase diagram at zero temperature, as reported in Fig. \ref{fig2}.
In our numerical calculations, using the number density $n$ we have
set the Fermi wavevector $k_{F}=(3\pi^{2}n)^{1/3}$ and Fermi energy
$E_{F}=\hbar^{2}k_{F}^{2}/(2m)$ as the units for wavevector and energy,
respectively. Unless specifically noted, we shall focus on the weak
coupling case with a dimensionless interaction parameter $1/(k_{F}a_{s})=-0.5$
and at zero temperature $T=0$, for which our mean-field treatment
could be well justified.

It is readily seen from the phase diagram that an in-plane Zeeman
field will drive the Fermi system from a gapped FF superfluid to a
gapless phase (labelled as ``nodal FF'') \cite{Dong2013NJP}. Remarkably,
at sufficiently large value it will also lead to a gapless topologically
non-trivial state (``topological nodal FF''). The evolution of the
energy spectrum at a typical spin-orbit coupling strength $\lambda=E_{F}/k_{F}$
as a result of the increasing in-plane Zeeman field is presented in
Fig. \ref{fig3}.

Physically, the transition to a gapless phase can be well characterized
by a global energy gap $E_{g}=\min E_{2+}(\mathbf{k})$, which is
the half of the energy difference between the minimum energy of the
particle branch and the maximum of the hole branch due to the particle-hole
symmetry $E_{2+}(\mathbf{k})=-E_{2-}(-\mathbf{k})$. Hence, $E_{g}\leq0$
and $E_{g}>0$ characterize a gapless and gapped state, respectively.
The topological phase transition, on the contrary, is related to the
change of the topology of the Fermi surfaces. It is well known that
such a change must be accompanied with the close and re-open of an
energy gap at some specific points in momentum space \cite{Hasan2010,Qi2011}.
In our continuum case of a homogeneous Fermi gas, this occurs precisely
at the origin $\mathbf{k}=\mathbf{0}$ (see Fig. \ref{fig3}b). Therefore,
naïvely the topological transition can be determined from the condition
$E_{2+}(\mathbf{k}=\mathbf{0})=0$ or more explicitly, 
\begin{equation}
h_{c2}=\sqrt{\left(\mu-\frac{\hbar^{2}q^{2}}{8m}\right)^{2}+\Delta^{2}}-\frac{\lambda q}{2}.
\end{equation}
In the absence of an FF pairing momentum ($q=0$), the above condition
reduces to the well-known criterion $h_{c}=\sqrt{\mu^{2}+\Delta^{2}}$
for the appearance of an SLTD topological superfluid when an out-of-plane
Zeeman field is applied \cite{Sau2010,Oreg2010}. It is interesting
that the gapless transition always occurs before the topological transition,
as a result of $E_{g}\leq E_{2+}(\mathbf{k}=\mathbf{0})$. Thus, bulk-gapped\emph{
}topological FF superfluids, if exist, must appear at very high in-plane
Zeeman field. As a superfluid analogue of strong 3D topological insulators
\cite{Hasan2010,Qi2011}, they are anticipated to have the unique
feature of a single Dirac cone for the energy dispersion of the Majorana
edge states. Unfortunately, in the parameter space that we considered,
we do not find their existence.

At the coupling strength $\lambda=E_{F}/k_{F}$, the gapless transition
and topological transition occur at $h_{c1}\simeq0.3E_{F}$ and $h_{c2}\simeq0.327E_{F}$,
respectively, as can be seen from Fig. \ref{fig3}a, where $E_{g}$
(red dashed line) and $E_{2+}(\mathbf{k}=\mathbf{0})$ (black solid
line) become zero as the in-plane Zeeman field increases. When $h>h_{c1}$,
nodal points which satisfy $E_{2\pm}(\mathbf{k})=0$ develop and form
two closed surfaces in momentum space \cite{Dong2013NJP}. When the
in-plane field further increases, passing through the threshold $h_{c2}$
for the topological transition (see Fig. \ref{fig3}d), the energy
dispersions of the particle- and hole-branches touch at two specific
points ($\pm k_{W}$,0,0), as shown in Fig. \ref{fig3}e. Around these
points, the dispersion of Bogoliubov quasiparticles in the bulk acquires
a linear structure and thereby form a Dirac cone. This is precisely
the energy dispersion for massless Weyl fermions \cite{Wan2011,Burkov2011,Xu2011}.
In this sense, the gapless topological FF superfluid predicted in
this work provides a new avenue for the observation of Weyl fermions
around the Weyl nodes ($\pm k_{W}$,0,0). Indeed, Weyl fermions have
recently been discussed in the context of 3D gapped topological superfluids
\cite{Sau2012,Xu2014}.

In our case, the appearance of the Weyl nodes and of the topological
order is closely related. Due to the asymmetry in the $k_{x}$ axis,
only one of the Weyl nodes is occupied. Thus, we may characterize
the topological order of the gapless FF superfluid by using the topological
invariant of Weyl fermions \cite{Qi2011,Schnyder2008}: 
\begin{equation}
N_{W}=\int\frac{d^{3}\mathbf{k}}{24\pi^{2}}\epsilon^{\mu\nu\rho}\textrm{Tr}\left[Q_{\mathbf{k}}^{\dagger}\partial_{\mu}Q_{\mathbf{k}}Q_{\mathbf{k}}^{\dagger}\partial_{\nu}Q_{\mathbf{k}}Q_{\mathbf{k}}^{\dagger}\partial_{\rho}Q_{\mathbf{k}}\right],
\end{equation}
where $Q_{\mathbf{k}}$ is the unitary matrix determined by the BdG
Hamiltonian, $\mu,\nu,\rho=(k_{x},k_{y},k_{z})$ and the domain of
the integration includes the isolated, occupied Weyl node. The gapless
topological FF superfluid is uniquely characterized a nonzero topological
invariant $N_{W}\neq0$.

To further demonstrate the topological nature of the gapless FF superfluid,
we calculate the energy dispersion in the presence of open boundary
by imposing cylindrical hard wall confinement perpendicular to the
\textit{y-z} plane with the $x$-axis being the symmetry axis (see
Methods). There is a pair of zero-energy Majorana fermion state on
the boundary of $r=\sqrt{y^{2}+z^{2}}=L$, which is the direct signature
of a topologically non-trivial state. The existence of Majorana fermions
is reported in Fig. \ref{fig3}a by green solid circles. They immediately
appear after the change in the topology of the Fermi surfaces.

\begin{figure}
\begin{centering}
\includegraphics[clip,width=0.48\textwidth]{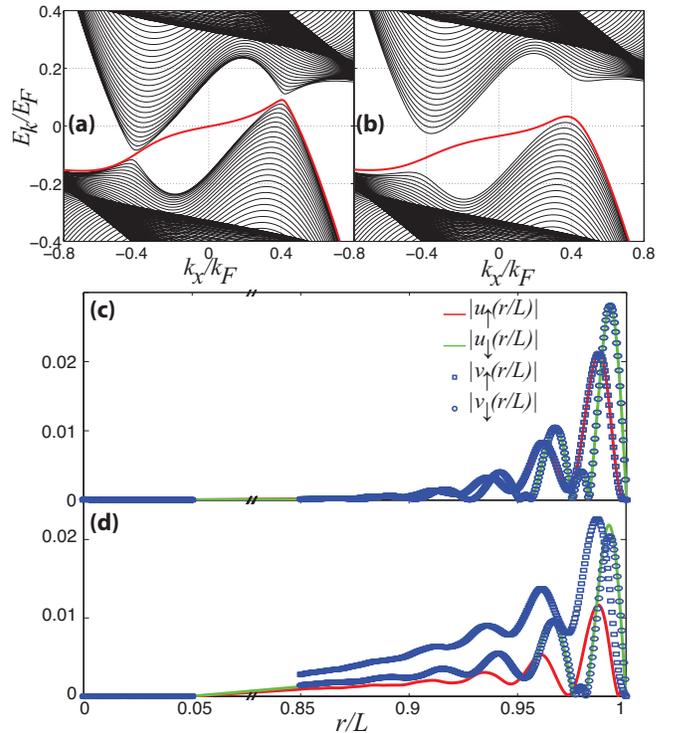} 
\par\end{centering}

\caption{(Color online) \textbf{Majorana surface states arising from the hard
wall confinement perpendicular to the }\textbf{\textit{y-z}}\textbf{
plane}. (\textbf{a}) and (\textbf{b}) The energy spectrum $E_{k_{x}}^{(m)}$
as a function of $k_{x}$ for $m=0$ and $m=10$, respectively. (\textbf{c})
The wavefunction of the zero-energy Majorana fermions for $m=0$,
satisfying the symmetry $u_{\sigma}(r)=e^{i\vartheta}v_{\sigma}^{*}(r)$,
where $\vartheta$ is a constant phase factor and $\sigma=\uparrow,\downarrow$.
(\textbf{d}) The wavefunction of the zero-energy surface state for
$m=10$. In numerical calculations, we have set the radius of the
confinement $L=200k_{F}^{-1}$. Other parameters are $\lambda=E_{F}/k_{F}$
and $h=0.4E_{F}$.}

\label{fig4} 
\end{figure}

We now discuss in more detail the Majorana surface states, whose dispersion
is shown in Fig. \ref{fig4}. Because the boundary we impose has cylindrical
symmetry in the \textit{y-z} plane and translational symmetry along
the \textit{x}-direction, the quasiparticle wavefunction takes the
following form: 
\begin{equation}
\Phi_{k_{x}\eta}^{\nu}=e^{ik_{x}x}e^{im\phi}\left[u_{\eta\uparrow}^{\nu}(r),u_{\eta\downarrow}^{\nu}(r)e^{i\phi},v_{\eta\uparrow}^{\nu}(r)e^{i\phi},v_{\eta\downarrow}^{\nu}(r)\right]^{T},
\end{equation}
where $(x,r,\phi)$ form the cylindrical coordinates. States with
different orbital angular momentum quantum number $m$ and linear
momentum $k_{x}$ are decoupled, see Methods for basis expansion.
In Figs.~\ref{fig4}a and \ref{fig4}b, we plot the energy spectrum
along $k_{x}$ axis for $m=0$ and $m=10$, respectively. Majorana
zero energy mode can be identified by the energy crossing of surface
state contribution, at which points quasiparticle wavefunctions become
localized near the boundary $r=L$, as shown in Figs.~\ref{fig4}c
and \ref{fig4}d. The surface states smoothly connect to the Weyl
nodes located approximately at $k_{W}\simeq0.4k_{F}$ in the bulk.
As $|m|$ increases, the localization of the zero-energy surface modes
deteriorates due to their hybridization with the bulk modes, and the
desired symmetry $u_{\sigma}(r)=e^{i\vartheta}v_{\sigma}^{*}(r)$
($\sigma=\uparrow$, $\downarrow$) for Majorana mode \cite{Liu2012}
is violated (see Fig.~\ref{fig4}d). One can prove that, as a result
of particle-hole symmetry of the BdG equation, when $m\rightarrow-m-1$,
and $k_{x}\rightarrow-k_{x}$, we have $E_{\eta\nu}\rightarrow-E_{\eta\nu}$.
Hence for every zero-energy state at given $m$ and $k_{x}$, there
is a corresponding zero-energy state at $-m-1$ and $-k_{x}$, which
describes the same physical state. As a result, it is easy to see
from Figs.~\ref{fig4}a and \ref{fig4}b that for arbitrary azimuthal
angular momentum $m$, the Majorana surface states have nearly the
same unidirected velocity $v(k_{x})=\partial E_{k_{x}}^{(m)}/\partial k_{x}>0$.
As there are no net atomic currents at equilibrium, the current carried
by these co-propagating surface states must be compensated by the
current induce by some extra counter-propagating modes in the bulk.
This is only possible when the system is gapless in the bulk, consistent
with the gapless nature of our topological FF superfluid. The unidirected
surface states discussed in our work is therefore a unique feature
of novel gapless topological FF superfluid. 

We may also consider imposing hard wall confinement along a specific
direction, for example, along the \textit{y}-direction at $y=0,L$.
In this case, unidirected Majorana surface states propagate in the
same direction on opposite boundaries at $y=0$ and $y=L$, respectively.
For detailed discussions, we refer to Supplementary Information.

It is worth noting that the 3D gapless topological FF superfluid can
not be viewed as a stack of 2D topological superfluids along a specific
direction (i.e., \textit{z}-axis), unlike the standard 3D topological
superfluids known so far. For the latter, the Majorana surface states
of the 3D system can be understood as the edge states of the 2D system
on the surfaces which are parallel to the \textit{z}-axis and therefore
have a flat dispersion that does not depend on $k_{z}$ \cite{Sau2012}.
This is analogous to the trivial or weak 3D topological insulators
\cite{Hasan2010,Qi2011}. In our case, due to the existence of spin-orbit
coupling in all three spatial directions, the dispersion of the Majorana
surface states is no longer flat. In this respect, the gapless topological
FF superfluid is better regarded as the superfluid analogue of a strong
topological insulator \cite{Hasan2010,Qi2011}, although the surface
states may not have a Dirac-cone-like dispersion due to the gapless
bulk.

\textbf{\textit{Discussions}}. --- Owing to their significant parameter
space in the phase diagram, our proposed gapless topological FF superfluids
could be easily detected in current cold-atom experiments, where the
isotropic or Rashba-type spin-orbit coupling can be engineered by
using Raman lasers \cite{Anderson2012} or a sequence of pulsed inhomogeneous
magnetic fields \cite{Anderson2013}. The unidirected Majorana surface
modes, as a unique experimental evidence of the novel topological
superfluid, could in principle be directly visible as arcs in momentum
space at the Fermi surfaces in the momentum-resolved radio-frequency
spectroscopy \cite{Liu2012}. In solid-state systems, a promising
candidate for realizing the gapless topological FF superfluid is to
use a quantum well with large Rashba and Dresselhaus spin-orbit couplings
(i.e., hole-doped InSb) in proximity to a conventional \textit{s}-wave
superconductor, where the in-plane Zeeman field in the quantum well
layer can be controlled to minimize the orbital effect \cite{Alicea2010}.
Other candidates include noncentrosymmetric superconductors such as
CePt$_{3}$Si and Li$_{2}$Pd$_{x}$Pt$_{3-x}$B \cite{Schnyder2011},
in the presence of a magnetic field on the plane of Rashba spin-orbit
coupling.

In summary, we have proposed a new mechanism to create topologically
non-trivial states by using an in-plane Zeeman field only. Novel gapless
topological superfluids with inhomogeneous Fulde-Ferrell pairing order
parameter can be realized using three-dimensional spin-orbit coupled
\textit{s}-wave superfluids, where the finite momentum pairing and
topological order are both driven by the in-plane Zeeman field. They
feature unidirected Majorana surface modes and a pair of zero-energy
Majorana fermions at the edges, which are quite different from the
standard gapped topological superfluids that are known to date. These
new features will greatly enrich our understanding of topological
quantum matters, in both solid-state and cold-atom systems.

\section*{Methods}

\textbf{\textit{Mean field theory}}. --- The details of our theoretical
framework have been presented in the previous work \cite{Dong2013NJP}.
Here we give a brief summary. Taking the mean-field approximation
for the pairing interaction term, the model Hamiltonian of the Fermi
system can be rewritten into a compact form, $\mathcal{H}=(1/2)\int d{\bf {\bf r}}\Phi^{\dagger}(\mathbf{{\bf r}})\mathcal{H}_{BdG}(\mathbf{\hat{k}})\Phi(\mathbf{{\bf r}})-V\Delta^{2}/U_{0}+\sum_{\mathbf{k}}(\xi_{\mathbf{k}+\mathbf{Q}/2}+\xi_{\mathbf{k}-\mathbf{Q}/2})/2$,
where the explicit form of $\mathcal{H}_{BdG}(\mathbf{\hat{k}})$
in Eq. (\ref{eq:BdGHami}) is given by

\begin{equation}
\left[\begin{array}{cccc}
\hat{\xi}_{{\bf k}+}+\lambda\hat{k}_{z} & \Lambda_{{\bf k}+}^{\dagger} & 0 & -\Delta\\
\Lambda_{{\bf k+}} & \hat{\xi}_{{\bf k}+}-\lambda\hat{k}_{z} & \Delta & 0\\
0 & \Delta & -\hat{\xi}_{{\bf k}-}+\lambda\hat{k}_{z} & \Lambda_{{\bf k-}}\\
-\Delta & 0 & \Lambda_{{\bf k-}}^{\dagger} & -\hat{\xi}_{{\bf k}-}-\lambda\hat{k}_{z}
\end{array}\right]\label{eq:BdGHami2}
\end{equation}
with $\hat{\xi}_{{\bf k}\pm}\equiv\hbar^{2}(\hat{\mathbf{k}}\pm\mathbf{Q}/2)^{2}/(2m)-\mu$
and $\Lambda_{{\bf k\pm}}\equiv\lambda(\hat{k}_{x}\pm q/2+i\hat{k}_{y})\pm h$.
For a homogeneous Fermi gas with open boundary condition, the BdG
Hamiltonian can be diagonalized by replacing the momentum operators
$\hat{k}_{i}$ ($i=x,y,z$) by the corresponding \textit{c}-numbers
$k_{i}$. Thus, we obtain the energy spectrum of Bogoliubov quasiparticles
$E_{\eta\nu}(\mathbf{k})$, where $\nu\in(+,-)$ denotes the particle
or hole branch and $\text{\ensuremath{\eta\in}(1,2)}$ stands for
the upper or lower helicity band. The mean-field thermodynamic potential
$\Omega_{\textrm{mf}}$ at the temperature $T$ can be written down
straightforwardly, 
\begin{eqnarray}
\frac{\Omega_{\textrm{mf}}}{V} & = & \frac{1}{2V}\sum_{\mathbf{k}}\left[\xi_{\mathbf{k}+\mathbf{Q}/2}+\xi_{\mathbf{k}-\mathbf{Q}/2}-\sum_{\eta=1,2}E_{\eta+}(\mathbf{k})\right]\nonumber \\
 &  & -\frac{\Delta^{2}}{U_{0}}-\frac{k_{B}T}{V}\sum_{\mathbf{k}\eta=1,2}\ln\left[1+e^{-E_{\eta+}(\mathbf{k})/k_{B}T}\right],
\end{eqnarray}
where the last term is the standard expression of thermodynamic potential
for non-interacting Bogoliubov quasiparticles. Due to the inherent
particle-hole symmetry in the Nambu spinor representation, the summation
over the quasiparticle energy has been restricted to the particle
branch, to avoid double counting. For a given set of parameters (i.e.,
the temperature $T$, \textit{s}-wave scattering length $a_{s}$ etc.),
different mean-field phases can be determined using the self-consistent
stationary conditions: $\partial\Omega_{\textrm{mf}}/\partial\Delta=0$,
$\partial\Omega_{\textrm{mf}}/\partial q=0$, as well as the conservation
of total atom number, $n=-(1/V)\partial\Omega/\partial\mu$, where
$n$ is the number density. At a given temperature, the ground state
has the lowest free energy $F=\Omega+\mu N$. For simplicity, we only
report the results at zero temperature.

\textbf{\textit{Majorana surface modes}}. --- To determine the Majorana
surface states in the topologically non-trivial phase, we impose a
cylindrical hard wall potential, for example, perpendicular to the
\textit{y-z} plane, so that any single-particle wavefunction must
vanish identically at the boundary $r=L$. We assume that the radius
is sufficiently large so that we can use the solution of a uniform
pairing gap. Accordingly, in the BdG Hamiltonian Eq. (\ref{eq:BdGHami2}),
we replace the momentum operator $k_{y}$ and $k_{z}$ with its corresponding
derivatives in cylindrical coordinates where longitudinal axis is
chosen along \textit{x}-direction. It can be diagonalized by using
the following ansatz for the Bogoliubov wavefunctions, 
\begin{equation}
\left[\begin{array}{c}
u_{\uparrow}({\bf r})\\
u_{\downarrow}({\bf r})\\
v_{\uparrow}({\bf r})\\
v_{\downarrow}({\bf r})
\end{array}\right]=\frac{e^{im\theta}}{\sqrt{2\pi}}\sum_{n=1}^{N_{\textrm{max}}}\left[\begin{array}{c}
\frac{J_{m}(\kappa_{n}^{(m)}r/L)}{\sqrt{\mathcal{N}_{n}^{(m)}}}u_{n\uparrow}\\
\frac{J_{m+1}(\kappa_{n}^{(m+1)}r/L)e^{i\theta}}{\sqrt{\mathcal{N}_{n}^{(m+1)}}}u_{n\downarrow}\\
\frac{J_{m+1}(\kappa_{n}^{(m+1)}r/L)e^{i\theta}}{\sqrt{\mathcal{N}_{n}^{(m+1)}}}v_{n\uparrow}\\
\frac{J_{m}(\kappa_{n}^{(m)}r/L)}{\sqrt{\mathcal{N}_{n}^{(m)}}}v_{n\downarrow}
\end{array}\right]e^{ik_{x}x}\label{eq:basisBessel}
\end{equation}
where $\kappa_{n}^{(m)}$ is the $n$th positive root of Bessel function
of the first kind $J_{m}(\rho)$ with $m\geq0$. For states with $m<0$,
we have instead $J_{-m}(\rho)=(-1)^{m}J_{m}(\rho)$. Orthogonal condition
is given by $\int_{0}^{L}J_{m}(\kappa_{n}^{(m)}r/L)J_{m}(\kappa_{l}^{(m)}r/L)rdr=0$
where integer $n\neq l$ and normalization reads as $\mathcal{N}_{n}^{(m)}=\int_{0}^{L}J_{m}(\kappa_{n}^{(m)}r/L)J_{m}(\kappa_{n}^{(m)}r/L)rdr=\frac{1}{2}L^{2}[J_{m+1}(\kappa_{n}^{(m)})]^{2}$.
Inserting this ansatz into the BdG equation, we convert the BdG Hamiltonian
into a $4N_{\textrm{max}}$ by $4N_{\textrm{max}}$ Hermitian matrix,
\begin{eqnarray}
\mathcal{H}_{11n}^{(m)}u_{n\uparrow}-i\lambda\sum_{l=1}^{N_{\textrm{max}}}\mathcal{W}_{ln}^{(m)}u_{l\downarrow}-\Delta v_{n\downarrow} & = & E_{k_{x}}^{(m)}u_{n\uparrow}\label{eq:bdg1}\\
i\lambda\sum_{l=1}^{N_{\textrm{max}}}\mathcal{W}_{nl}^{(m)}u_{l\uparrow}+\mathcal{H}_{22n}^{(m)}u_{n\downarrow}+\Delta v_{n\uparrow} & = & E_{k_{x}}^{(m)}u_{n\downarrow}\label{eq:bdg2}\\
\Delta u_{n\downarrow}+\mathcal{H}_{33n}^{(m)}v_{n\uparrow}+i\lambda\sum_{l=1}^{N_{\textrm{max}}}\mathcal{W}_{nl}^{(m)}v_{l\downarrow} & = & E_{k_{x}}^{(m)}v_{n\uparrow}\label{eq:bdg3}\\
-\Delta u_{n\uparrow}-i\lambda\sum_{l=1}^{N_{\textrm{max}}}\mathcal{W}_{ln}^{(m)}v_{l\uparrow}+\mathcal{H}_{44n}^{(m)}v_{n\downarrow} & = & E_{k_{x}}^{(m)}v_{n\downarrow}\label{eq:bdg4}
\end{eqnarray}
where all the matrix elements have been analytically worked out (not
shown here). The diagonalization directly gives rise to the energies
and wavefunctions of the Majorana surface states.

\paragraph*{\textbf{Acknowledgement}}

HH and XJL were supported by the ARC Discovery Projects (Grant Nos.
FT130100815, DP140103231 and DP140100637) and NFRP-China (Grant No.
2011CB921502). HP was supported by the NSF, the Welch Foundation (Grant
No. C-1669) and the DARPA OLE program.

\paragraph*{\textbf{Author contributions}}

HH, HP and XJL concieved the concept of topolgoical superfluidity
induced by an in-plane Zeeman field. HH, LD and XJL performed numerical
calculations. YC checked the topological superfluidity in the case
of two-dimensional atomic Fermi gases. All authors contributed to
the writting of the manuscript. Correspondence and requests for numerical
data should be addressed to X.-J. Liu (email: xiajiliu@swin.edu.au).

\paragraph*{\textbf{Competing financial interests}}

The authors declare no competing finanical interests.

\begin{figure*}
\begin{centering}
\includegraphics[clip,width=0.7\textwidth]{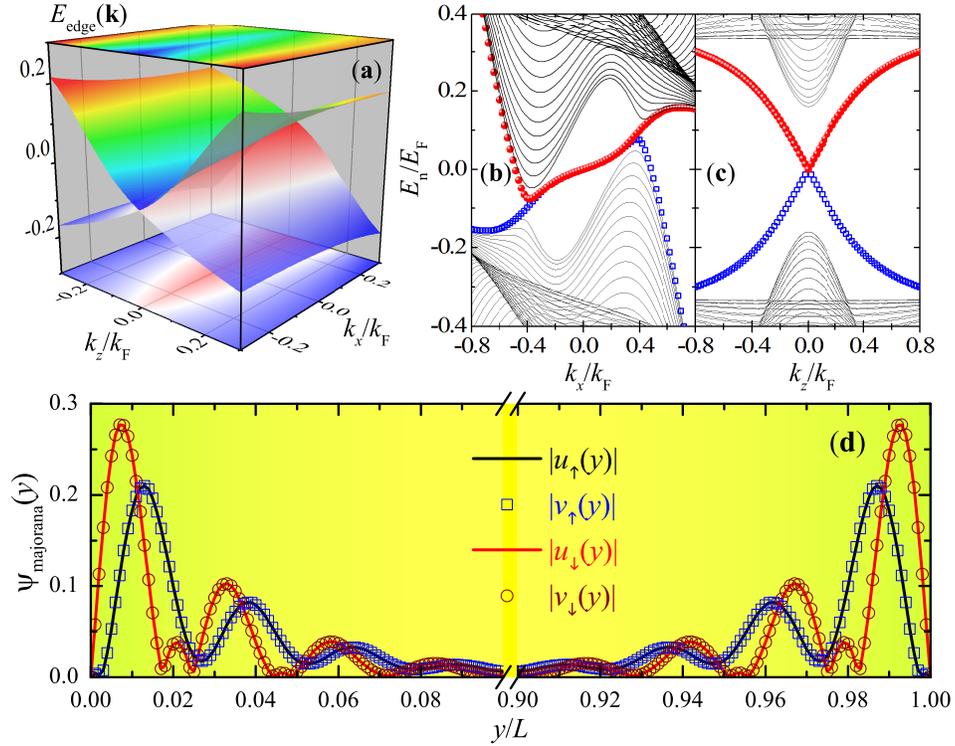} 
\par\end{centering}

\caption{(Color online) \textbf{Majorana surface states arising from the hard
wall confinement along the }\textbf{\textit{y}}\textbf{-direction}.
(\textbf{a}) The surface state dispersion forms two sheets which crosses
at the line $k_{z}=0$. (\textbf{b}) and (\textbf{c}) The full energy
spectrum $E_{2\pm}(k_{x},k_{z})$ along the $k_{z}$ or $k_{x}$ axis,
respectively. The surface states at the two boundaries are highlighted
by red solid circles and blue empty squares, respectively. (\textbf{d})
The wavefunction of the zero-energy Majorana fermions at $k_{x}=0$
and $k_{z}=0$. In numerical calculations, we have set the length
of the confinement $L=200k_{F}^{-1}$. Other parameters are $\lambda=E_{F}/k_{F}$
and $h=0.4E_{F}$ as in Fig. \ref{fig4}.}

\label{figS1} 
\end{figure*}

\section*{Supplementary information}

Here we discuss the Majorana surface states with a hard wall potential
along a specific direction, say, along the \textit{y}-direction. Any
single-particle wavefunction must vanish identically at the boundary
$y=0$ or $y=L$. We assume that the length $L$ is sufficiently large
so we use the solution of a uniform pairing gap. Accordingly, in the
BdG Hamiltonian Eq. (\ref{eq:BdGHami2}), we replace the momentum
operator $k_{y}$ with $-i\partial_{y}$. It can be diagonalized by
using the following ansatz for the Bogoliubov wavefunctions,
\begin{eqnarray}
u_{\sigma}\left(y\right) & = & \sum_{n=1}^{N_{\textrm{max}}}u_{n\sigma}\psi_{n}\left(y\right),\\
v_{\sigma}\left(y\right) & = & \sum_{n=1}^{N_{\mathbf{\textrm{max}}}}v_{n\sigma}\psi_{n}\left(y\right),
\end{eqnarray}
where $\psi_{n}\left(y\right)=\sqrt{2/L}\sin[n\pi y/L]$ is the eigenfunction
of the hard wall potential with eigenvalue $\epsilon_{n}=\hbar^{2}n^{2}\pi^{2}/(2mL^{2}).$
Inserting this ansatz into the BdG equation, we convert the BdG Hamiltonian
into a $4N_{\textrm{max}}$ by $4N_{\textrm{max}}$ symmetric matrix,
whose diagonalization directly leads to the energies and wavefunctions
of the Majorana surface states.

With this hard wall confinement, the dispersion of Majorana surface
states is shown in Fig. \ref{figS1}. In momentum space, $k_{x}$
and $k_{z}$ are still good quantum numbers, so we actually plot $\min E_{2+}(k_{x},k_{z})$
and $\max E_{2-}(k_{x},k_{z})$. There are two sheets in the energy
dispersion (Fig. \ref{figS1}a), corresponding to the surface states
localized at the boundary $y=0$ and $y=L$, respectively. Remarkably,
these two sheets cross at the line $k_{z}=0$, indicating that along
this line the two branches of surface states are unidirected, that
is, propagating in the same direction on opposite boundaries. This
is highlighted in Fig. \ref{figS1}b, from which we also identify
that the unidirected Majorana surface states smoothly connect the
two Weyl nodes ($\pm k_{W}$,0,0) in the bulk, where $k_{W}\simeq0.4k_{F}$.
Recall that at equilibrium there are no net atomic currents. As in
the cylinderically symmetric case, the current due to these co-propagating
surface states on opposite boundaries therefore must be compensated
by the current induced by some extra counter-propagating modes in
the bulk. This can only happen in systems with a gapless bulk. We
note that, the unidirected Majorana surface states only occur along
the line $k_{z}=0$. Actually, if we make a cut on the two sheets
along other directions, for example, along the line $k_{x}=0$, it
is easy to see that the Majorana surface states become counter-propagating
(see Fig. \ref{figS1}c for the dispersion as a function of $k_{z}$),
resembling the surface states in the standard gapped topological superfluid.
This follows the fact that at $k_{x}=0$ our topological FF superfluid
is actually gapped in the bulk. By comparing the two limiting cases
shown in Figs. \ref{figS1}b and \ref{figS1}c, it is clear that the
unidirected surface states discussed in our work is a unique feature
of novel gapless topological FF superfluid.
\end{document}